\documentclass[reprint,superscriptaddress,amsmath,amssymb,aps]{revtex4-2}

\usepackage{bm}
\usepackage{float}
\usepackage{graphicx}
\usepackage[usenames,dvipsnames]{color}
\usepackage[normalem]{ulem}
\usepackage[svgnames]{xcolor}
\usepackage{multirow}
\usepackage{titlesec}
\usepackage[utf8]{inputenc}
\usepackage{ragged2e} 
\begin{document}

\title{Deviation from Fermi-liquid $T^2$ resistivity caused by collective transport}
\author{Kamran Behnia}
\affiliation{Laboratoire de Physique et d'\'Etude des Mat\'eriaux \\ 
(ESPCI - CNRS - Sorbonne Universit\'e), PSL Research University, 75005 Paris, France}
\date{\today}

\begin{abstract}
`Fermi-liquid behavior' is nowadays used as a shorthand for quadratic temperature dependence of the electrical resistivity. In a metal, for this to occur, electrons must be cooled well below the Fermi degeneracy temperature, $T_{\mathrm{F}}$. A departure from this behavior is observed at finite temperature. Here, we note that this departure has opposite signs in two distinct types of Fermi liquids. In weakly correlated ones, the upward departure implies a higher exponent of the inelastic resistivity, attributable to \textit{additional scattering} by phonons. In contrast, there is a downward deviation in strongly correlated metals, which implies either \textit{reduced scattering} or \textit{additional conduction}, as in the case of normal liquid $^3$He, in which a distinct contribution to heat transport by a sound mode has been recently identified. Such a channel of conduction smoothly transforms across $T_{\mathrm{F}}$ to the one captured by the Bridgman formula for thermal conductivity of classical liquids. In three metals (UPt$_3$, Sr$_2$RuO$_4$, and heavily overdoped LSCO), according to the available experimental data, the normalized amplitude of excess conductivity at $T/T_{\mathrm{F}}\simeq0.02$ is comparable to what has been seen in $^3$He, suggesting, without establishing, a common origin. 
\end{abstract}
\maketitle
\section{Introduction}
Soon after its formulation \cite{Landau1957}, Landau’s Fermi liquid theory became a cornerstone of twentieth-century condensed-matter physics. The original motivation for this theory was to explain the thermodynamic and transport properties of normal liquid $^3$He \cite{Abrikosov_1959}. Subsequently, and non-trivial obstacles notwithstanding, it was applied to electrons in metals \cite{Silin1959}. Despite the long-range Coulomb interaction, the absence of Galilean invariance and the presence of periodicity defects, the concept of the Landau quasiparticle survived in this territory and proved to be remarkably fertile. It was successfully employed to explain why in strongly correlated metals known as Kondo lattices or heavy fermion systems \cite{Stewart1984,Lee1986}, the electronic specific heat (the Sommerfeld coefficient) and the cyclotron mass detected by quantum oscillations \cite{Taillefer1988} are both amplified by an identical factor. The renormalized heavy mass of electrons amplifies all transport and thermodynamic coefficients, such as Pauli magnetic susceptibility \cite{Wilson1975}, the $T^2$ inelastic electrical resistivity \cite{Rice1968,KADOWAKI} and the $T$-linear diffusive Seebeck coefficient \cite{Behnia_2004}. 

The success of Landau's paradigm has shifted the focus of investigation to metals which appear to reside beyond it, nowadays called `non-Fermi liquid' or `strange' metals \cite{Phillips2022}. Nevertheless, even in well-established Fermi liquids, there remain a number of unresolved puzzles \cite{Behnia2022}.

One is the nature of the microscopic mechanism of dissipation caused by electron-electron scattering. The $T^2$ temperature dependence of the inelastic scattering reflects the growth of the phase space of scattering between quasiparticles \cite{Mott1990}. But given that momentum [as well as pseudo-momentum] is conserved, why the charge current is degraded by such scattering? If this question has ever been given much thought, its answer has been either Umklapp scattering events or a bottleneck caused by the multiplicity of electronic reservoirs. However, the persistence of $T^2$ resistivity in dilute metals with a single Fermi pocket located at the Brillouin zone center \cite{lin2015,Wang2020} suggests otherwise.

Another open question is the identity of the Fermi liquid's collective modes. Ironically, this was the subject of the first paper focusing on a metallic Fermi liquid \cite{Silin1959}. The most celebrated among them is zero sound, a collisionless collective mode, which, instead of being sustained by frequent collisions as in ordinary sound, is governed by interactions between quasiparticles acting as a self-consistent restoring force. Leggett \cite{Leggett_2016} states that the transverse zero sound is a unique feature of Landau's theory with no counterpart in a weakly interacting system. In $^3$He, zero sound has been experimentally observed \cite{Abel1966,Albergamo2007,Godfrin2012}. On the other hand, in metallic Fermi liquids, its presence, as well as its distinction from other collective modes such as plasmons \cite{Pines1952} or demons \cite{Husain2023}, is yet to be clarified. Any role it may play in the transport of charge and entropy remains an open question.

\begin{figure*}[ht!]
\centering
\includegraphics[width=1\linewidth]{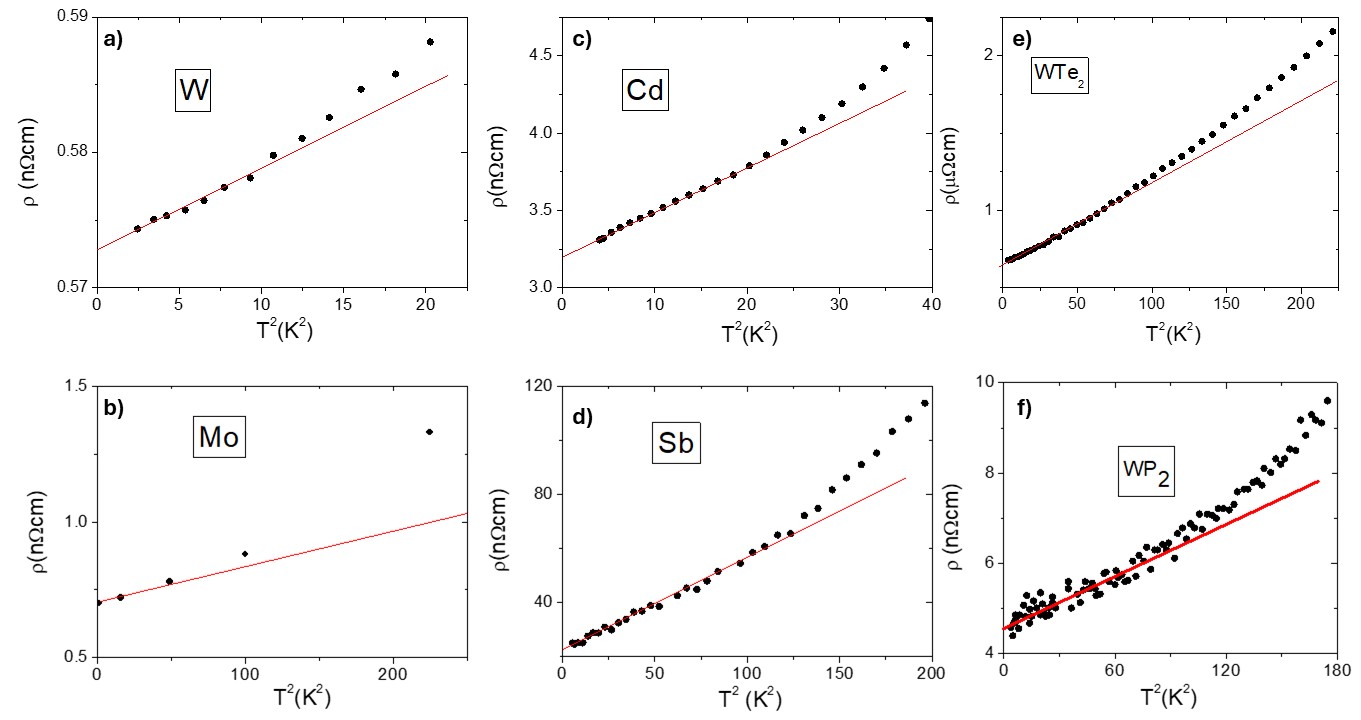} 
\caption{\textbf{Deviation from $T^2$ resistivity in weakly correlated Fermi liquids.} The electrical resistivity of W \cite{wagner1971} (a), Mo \cite{Desai1984} (b), Cd \cite{guo2026hydrodynamicsviscouselectronfluid} (c), Sb \cite{jaoui2021thermal} (d), WTe$_2$ \cite{Xie2024} (e) and WP$_2$ \cite{jaoui2018}(f) as a function of $T^2$. In all these cases, the data points deviate upward from the low-temperature $T^2$ behavior shown by the red line. This implies that the inelastic resistivity acquires an exponent larger than two, which is the behavior expected in the presence of additional scattering by phonons. }
\label{fig.Weak}
\end{figure*}

The subject matter of the present paper is a third puzzle with potential links to those mentioned above. According to the available experimental data in all hitherto-investigated metallic Fermi liquids, the canonical $T^2$ thermal resistivity is interrupted at a finite temperature that is orders of magnitude lower than the Fermi degeneracy temperature. By itself, this is not surprising. Scattering by phonons leads to a $T^5$ contribution to inelastic resistivity \cite{ziman1972}. This term is expected to arise in the so-called Bloch--Gr\"uneisen regime of electron--phonon scattering at temperatures well below the Debye temperature. Such a faster-growing term would eventually overwhelm the quadratic one. This picture holds in weakly correlated Fermi liquids (that is, those without significant mass amplification). According to the available data, in such metals, the deviation from $T^2$ resistivity is upward, indicating a passage from quadratic inelastic resistivity to a regime in which the relevant exponent is larger than two. In contrast to these well-understood cases, in many strongly correlated Fermi liquids, possibly in all of them, the deviation is downward. We will argue that this observation implies additional conduction instead of additional scattering. There appears to be a conducting channel on top of the quasiparticle channel, as in the case of $^3$He \cite{Behnia2024}. In the latter case, this additional conducting channel appears to be a hydrodynamic sound mode, with a roton-like minimum at $q=2k_F$ \cite{Dornheim2022,Filinov2023}. 


\section{Finite temperature deviation from $T^2$ resistivity in weakly correlated metals}

Figure \ref{fig.Weak} presents plots of the electrical resistivity as a function of $T^2$ in six different metals, according to the previously reported data. In all these metals, the effective mass of the electrons is equal to or smaller than the free-electron mass. In other words, electron correlations are too weak to generate any significant mass amplification. In all panels, the presence of a $\rho=\rho_0+AT^2$ dependence in the low-temperature limit is marked by a red solid line. In all cases, at some finite temperature, data points begin to deviate upward from this behavior. 

This upward deviation can be easily understood given that scattering of electrons by phonons gives rise to an additional term in the inelastic resistivity, which has a $T^5$ dependence for a three-dimensional metal, cooled well below its Debye temperature \cite{ziman1972}. The concentration of acoustic phonons in the Debye approximation follows $T^3$. This sets the temperature dependence of the electron--phonon scattering rate. In the case of electrical conductivity, there is an additional damping factor, due to the fact that small-angle scattering is not very efficient in decaying the momentum flux and the fraction of the phonon population capable of degrading momentum flux decreases quadratically (linearly) with temperature in 3D (2D). As a consequence, the expected temperature dependence is $T^4$ in two dimensions \cite{Efetov2010} and $T^5$ in three dimensions. 

The data shown in the figure are not exhaustive. The behavior shown in Figure \ref{fig.Weak} is generic to elemental metals, with the exception of noble metals in which Kondo impurities have hindered the observation of a clear $T^2$ behavior. Only four metallic elements (W, Mo, Cd, and Sb) are shown here, but there are many others. See, for example, Ref. \cite{Bass1990} for alkali metals, Ref. \cite{Garland_1978} for Al, and ref. \cite{White1959} for transition metals. The figure includes two metallic compounds (WP$_2$ and WTe$_2$). 

There are other weakly correlated metals, in which $T^2$ and $T^5$ terms in the temperature-dependent resistivity have both been quantified. Among cases attracting recent attention, one can mention RuO$_2$\cite{ling2026} (a metallic oxide) and CsV$_3$Sb$_5$ \cite{Menil2026} (a kagome metal and superconductor).

Let us now turn to strongly correlated metals. 

\begin{figure*}[ht!]
\centering
\includegraphics[width=1\linewidth]{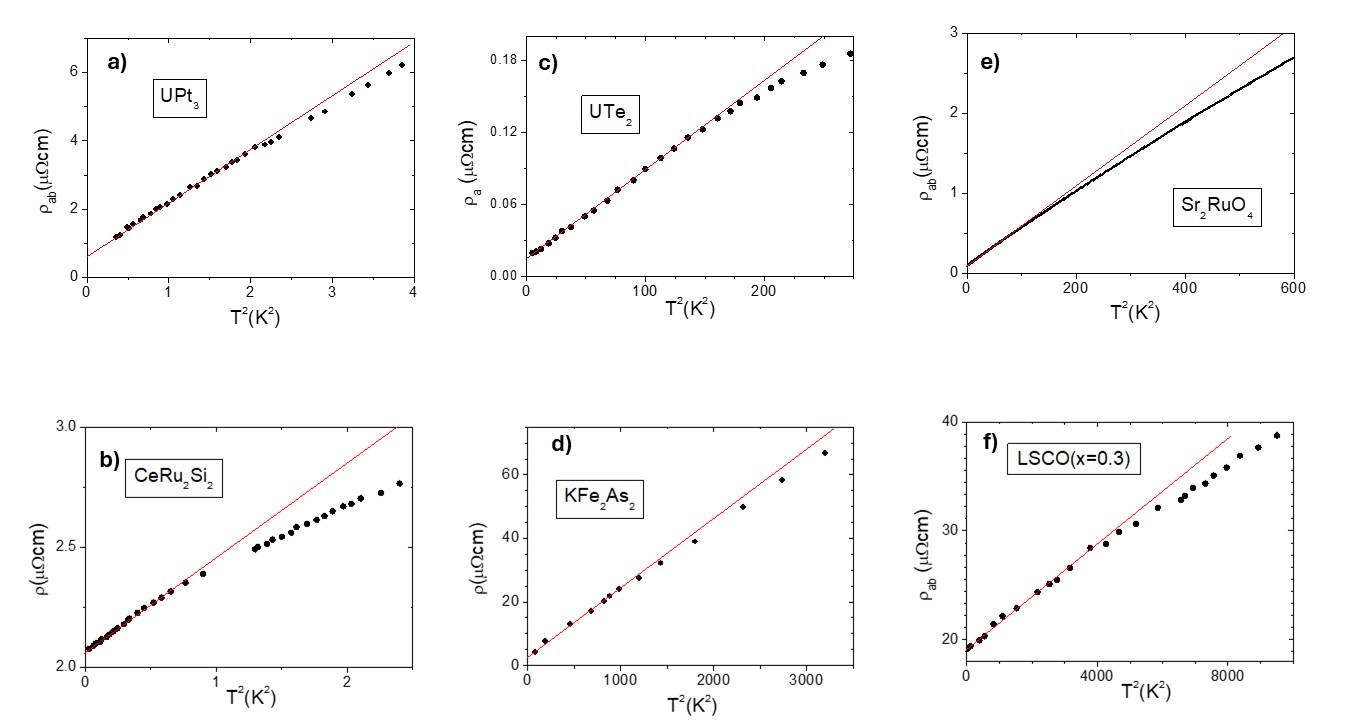} 
\caption{\textbf{Deviation from $T^2$ resistivity in strongly correlated Fermi liquids.} The electrical resistivity of UPt$_3$ \cite{Joynt2002}(a), CeRu$_2$Si$_2$ \cite{Mignot1989} (b), UTe$_2$ \cite{Eo2022} (c), KFe$_2$As$_2$ \cite{Hardy2013} (d), Sr$_2$RuO$_4$\cite{Barber2018} (e) and La$_{1.7}$Sr$_{0.3}$CuO$_4$\cite{Nakamae2003} (f) as a function of $T^2$. In these cases, the data points deviate downward from the low-temperature $T^2$ behavior shown by the red line. This implies the presence of an additional conducting channel.}
\label{fig.strong}
\end{figure*}
\section{Finite temperature deviation from $T^2$ resistivity in strongly correlated metals}

Figure \ref{fig.strong} shows electrical resistivity as a function of $T^2$ in six other metals. In contrast to those shown in Fig. \ref{fig.Weak}, these metals are strongly correlated. They host electrons whose mass is much larger than the free-electron mass. This time the deviation from quadratic temperature dependence is downward. 

The compounds selected for this figure come from a variety of families of materials with strong electronic correlations. UPt$_3$ \cite{Joynt2002} is a heavy-fermion superconductor ($T_c=0.5$ K), extensively explored for several decades. Its Sommerfeld coefficient is $\gamma=0.44\,\mathrm{J\,K^{-2}\,mol^{-1}}$, implying an amplification of the electron mass by more than two orders of magnitude. As seen in the figure, its resistivity displays a quadratic temperature dependence only below 1.5 K. The Fermi degeneracy temperature can be quantified from the oscillation frequency and the cyclotron mass of each Fermi surface pocket \cite{NORMAN1988245}. The resulting Fermi temperatures range from 29 K to 87 K \cite{lin2015}. 

CeRu$_2$Si$_2$ ($\gamma=0.35\,\mathrm{J\,K^{-2}\,mol^{-1}}$) is another heavy-fermion metal that does not undergo an ordering transition down to the lowest explored temperature \cite{FLOUQUET2002251}. It shows a quadratic temperature dependence when $T<0.9$ K. This is also well below its lowest Fermi energy ($\sim 40$ K), which can be extracted from quantum oscillations data \cite{AOki2014}.

UTe$_2$ \cite{Ran2019,Aoki_2022} is another unconventional superconductor with moderately heavy electrons ($\gamma=0.12\,\mathrm{J\,K^{-2}\,mol^{-1}}$). Its resistivity displays a $T^2$ temperature dependence along all three high-symmetry axes \cite{Aoki_2022,Eo2022}. Figure \ref{fig.strong}(c) shows the data along one crystalline axis. A downward deviation from the canonical $T^2$ behavior starts above 10 K. Combining the Sommerfeld coefficient with a carrier density of $1.1 \times 10^{28}$ cm$^{-3}$ yields a quasiparticle Fermi energy of $\sim$ 520 K, fifty times larger than the temperature at which the deviation begins.

KFe$_2$As$_2$ is an iron-based superconductor ($T_c\simeq3.5$ K) with $\gamma \simeq 0.1\,\mathrm{J\,K^{-2}\,mol^{-1}}$ \cite{Abdel2012}. As seen in Figure \ref{fig.strong}(d), its resistivity displays a $T^2$ temperature dependence up to $\approx$ 30 K, above which a downward deviation is visible. It has been proposed that a sizable fraction of $\gamma$ does not originate from mobile carriers \cite{grinenko2012}. We therefore refrain from quantifying the Fermi energy. However, it is undoubtedly orders of magnitude larger than 30 K. 

 Sr$_2$RuO$_4$ is a clean strongly correlated Fermi liquid with an unconventional superconducting ground state ($T_c=1.5$ K) and $\gamma \simeq 0.038\,\mathrm{J\,K^{-2}\,mol^{-1}}$ \cite{Maeno1997}. Its Fermi surface, known in great detail \cite{Mackenzie2003}, consists of three sheets, which are distorted two-dimensional cylinders. Their cyclotron masses and their quantum-oscillation frequencies yield Fermi temperatures ranging from $\sim$ 1200 K to $\sim$ 2400 K. Figure \ref{fig.strong}(e) shows its resistivity as a function of $T^2$, revealing a downward deviation starting at $T>15$ K. 
 
La$_{1.7}$Sr$_{0.3}$CuO$_4$ is a heavily doped cuprate in which superconductivity is suppressed by extensive doping. In the presence of the superconducting ground state (that is when $p<0.3$), there is a $T$-linear term in resistivity \cite{Cooper2009}. But as soon as it is destroyed, the linear term vanishes and the temperature dependence of resistivity corresponds to $\rho_0+AT^2$ \cite{Cooper2009}. Figure \ref{fig.strong}(f) shows its resistivity as a function of $T^2$, revealing a downward deviation starting at $T>40$ K. The Fermi temperature of the system is $\sim 5900$ K \cite{Jin2021}. 

\begin{figure*}[ht!]
\centering
\includegraphics[width=1\linewidth]{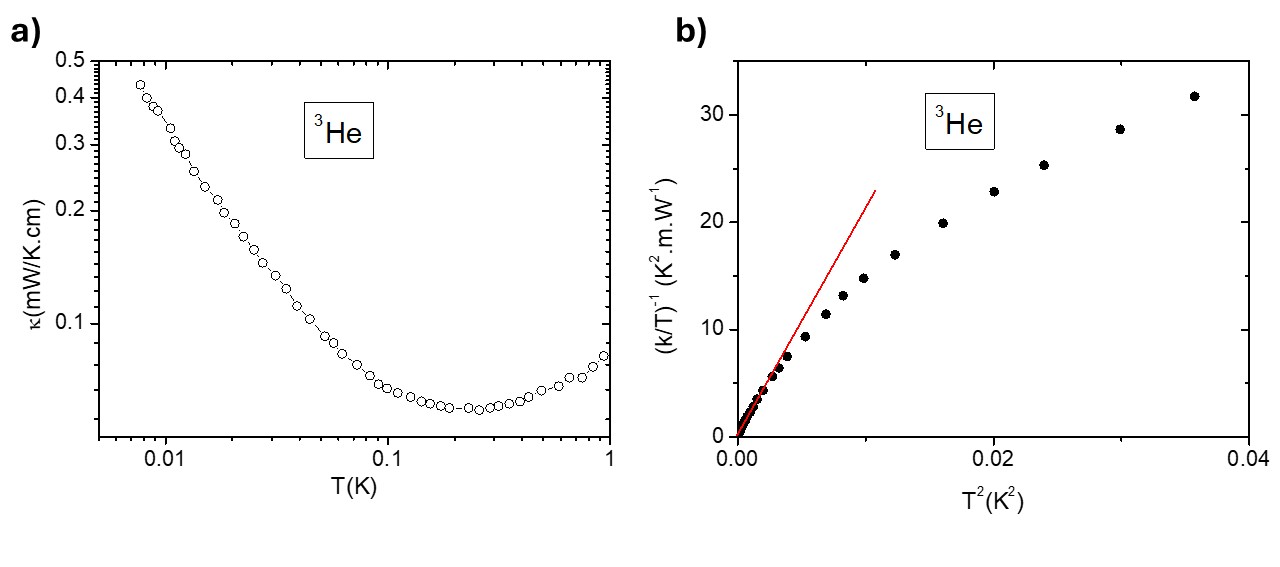} 
\caption{\textbf{Thermal conductivity and thermal resistivity in $^3$He} (a) Thermal conductivity, $\kappa$, of $^3$He as a function of temperature. (b) Thermal resistivity of $^3$He, defined as $(\kappa/T)^{-1}$, as a function of $T^2$ up to T=0.2 K. The data are identical to panel (a). The red solid line shows the low-temperature Fermi-liquid regime. Note the downward deviation at finite temperature.}
\label{fig.3He}
\end{figure*}

This figure does not represent an exhaustive list of metals exhibiting a similar behavior either. The downward deviation from the canonical $T^2$ temperature dependence highlighted here is generic to heavy-fermion compounds \cite{Stewart1984,Stewart2001}. In contrast to the behavior illustrated in Figure \ref{fig.Weak}, there is no established explanation for a downward deviation from the canonical behavior at a relatively low finite temperature in strongly correlated metals. 

Several authors have put forward the idea that this is caused by the loss of coherence and a concomitant decrease in mass amplification \cite{Burdin2000,Nakatsuji2004,Haule_2009,Mravlje2011}. This scenario implies the existence of a temperature scale, $T^*$ or $T_{\mathrm{coh}}$, significantly lower than the renormalized Fermi temperature. We will argue below that an alternative explanation exists, which does not require such a hypothesis. 

Before formulating it, we need to examine heat transport.

\section{Low-temperature heat transport in metallic and $^3$He Fermi liquids}

In the limit of elastic scattering, Fermi liquids obey the Wiedemann-Franz (WF) law. According to this law, in the absence of inelastic scattering, the ratio of $\kappa/T$, where $\kappa$ is the thermal conductivity, to the electrical conductivity, $\sigma$, is set by fundamental constants.

Therefore, the zero-temperature residual electrical resistivity, $\rho_0$, which is set by impurities, has a thermal counterpart: $WT_0\equiv(\frac{\kappa}{T})^{-1} (T \rightarrow 0)$, proportional to it by a constant (L$_0=2.44 \times 10^{-8}V^2/K^2)$.

As we saw in the previous section, inelastic scattering between electrons generates an enhancement in electrical resistivity, quadratic in temperature. This scattering gives rise to a similar enhancement in the thermal resistivity with a quadratic temperature dependence $WT\equiv(\frac{\kappa}{T})^{-1}$. The prefactors of the thermal and electrical resistivities are no longer strictly linked by $L_0$, which means that the WF law is no longer strict. However, it remains a valid approximation: not only the two resistivities do have the same temperature dependence, but also their ratio remains of the order of $L_0$. Even if there is no strict equality, the temperature dependence of the scattering phase space is the same.

Experimental studies have resolved a $T^2$ contribution to the thermal resistivity in several weakly correlated \cite{white1970,wagner1971,jaoui2018,jaoui2021thermal,Jiang2023,Gourgout2024,ling2026} and strongly correlated \cite{Lussier1994,paglione2005,Hassinger2017} Fermi liquids (see \cite{Behnia2022} for a review). The prefactor of the quadratic thermal resistivity has always been found to be larger than its electrical counterpart, a feature expected due to the relative immunity of electrical transport to small-angle scattering \cite{ziman1972}. Note, however, that the \textit{order of magnitude} respects the WF law. In elemental tungsten \cite{wagner1971}, which shows the largest credible discrepancy, the thermal prefactor exceeds its electrical counterpart by a factor of seven.

Normal liquid $^3$He was the original inspiration of Landau's Fermi liquid theory. In the second half of the twentieth century, its thermal conductivity was a focus of theoretical \cite{Abrikosov_1959, Nozieres, Brooker1968,Sykes1970,wolfle1979} and experimental \cite{Wheatley,abel1967,greywall1984, Murphy1994} attention. Both theory and experiment agreed that when $T \rightarrow 0$, $\kappa \propto T^{-1}$. 

The origin of this temperature dependence is identical to the $T^2$ thermal resistivity of metallic Fermi liquids. Here, in contrast to solids, disorder does not play any detectable role. With increasing temperature, the phase space for fermion-fermion scattering grows quadratically with temperature. As a result, the thermal diffusivity $D_{th}$ decreases quadratically with temperature. Given that the specific heat of degenerate fermions is linear in temperature, one would expect the thermal conductivity $\kappa \propto D_{th}C_V$ to follow $T^{-1}$. In a first approximation, this is what theory expects \cite{Abrikosov_1959, Nozieres, Brooker1968,Sykes1970,wolfle1979,Dobbs} and what experiments find \cite{Wheatley,abel1967,greywall1984, Murphy1994}. 

The two statements regarding the temperature dependence of thermal conductivity in Fermi liquids ($\kappa_{FL} \propto T^{-1}$ in $^3$He and $WT_{FL} \propto T^{2}$ in metals) are strictly equivalent. Nevertheless, a quantitative comparison between the two cases was not made before 2018 \cite{jaoui2018}. Such a comparison reveals a smooth continuity between $^3$He and strongly correlated metallic Fermi liquids. Fermi liquid parameters, energies and wave-vectors set the amplitude of the quadratic prefactor, providing a quantitative clue to the correlation first noticed by Kadowaki and Woods \cite{KADOWAKI} (following Rice \cite{Rice1968}) and recently extended to dilute metals, in which the standard Kadowaki-Woods scaling does not hold \cite{Wang2020} [for a more detailed discussion of this point see \cite{Behnia2022,Gourgout2024}]. 

Figure \ref{fig.3He} shows the temperature dependence of the thermal conductivity in $^3$He according to Greywall's data \cite{greywall1984}. This is the most detailed set of reported data and is in reasonable agreement with earlier measurements by Wheatley \cite{Wheatley} and Abel \cite{abel1967}. Panel (a) shows the data as originally plotted \cite{greywall1983}. There is a $\propto T^{-1}$ behavior at the lowest temperatures. But a deviation from this behavior appears at finite temperature, and above $\approx 0.25$ K, the thermal conductivity begins to increase with temperature. In other words, the `Fermi-liquid behavior' in $^3$He is restricted to very low temperature ($T<0.2$ K). Panel (b) shows the same data, but this time, the quantity plotted is $WT\equiv (\kappa/T)^{-1}$ \textit{vs.} $T^2$. One can see that at very low temperature, there is a quadratic temperature dependence followed by a downward deviation, very similar to what is seen in the panels of Figure \ref{fig.strong}.

In the case of $^3$He, the observation does not leave any possibility to escape an obvious conclusion. This is a Fermi liquid without disorder, lattice, or any distinction whatever between Umklapp and Normal events. Another channel of conduction when $T>0.01T_F$ is indisputably present \cite{Behnia2024}. 

\section{Collective transport in classical liquids}
In a liquid, in contrast to a gas, phonons play a key role in heat transport. In 1923, Bridgman \cite{Bridgman1923} derived a very simple equation for the thermal conductivity of liquids. This equation can be written in this concise way:
\begin{equation}
\kappa = r k_B v_s n^{2/3}
\label{Bridgman}
\end{equation}

Here, $n$ is the particle density and $v_s$ is the sound velocity. \(r\) is a dimensionless parameter, usually, but not always, taken as $r\simeq2$. 

A hundred years after its formulation, this equation remains influential and relevant \cite{Ohtori2014,KHRAPAK2023121786,Xi_2020,Zhao2021}, not only because of its successful account of the experimental data, but also because of its conceptual simplicity. 

It can be derived by invoking the kinetic formula. Thermal conductivity is set by the specific heat per volume of heat carriers, multiplied by their velocity and mean free path. Assuming that an entropy of $\sim r k_B$ per atom is traveling with the speed of sound over an interatomic distance leads to Eq.~\ref{Bridgman}. 

Deriving collective wave propagation of a temperature wave coupled to sound is beyond the relaxation time approximation of the Boltzmann equation. It requires its hydrodynamic limit \cite{Golse}. Starting from the three Euler equations, a wave equation for temperature, and not the usual diffusive heat equation, can be derived under isentropic conditions. The Bridgman equation emerges upon the assumption that the temperature wave attenuates over a distance $a$. 

The presence of $v_s$ in Eq.~\ref{Bridgman} implies that heat conduction in liquids is collective rather than single-particle diffusive. This is yet another example of the relevance of the concept of phonons, originally developed for crystalline solids, to the liquid state. The phonon theory of solids \cite{Trachenko_2016,trachenko2023theory} has significantly advanced our understanding of liquid thermodynamics and transport properties. Its conceptual accomplishment consists in providing a unified framework for all states of matter. Instead of treating liquids as a completely unique state, the phonon theory treats liquids as disordered media in which the high-frequency vibrational spectrum smoothly evolves as a function of temperature and pressure. This effectively bridges the gap between the elasticity of solids and the kinetic diffusion of gases.

\section{Collective transport in quantum liquids}
As seen in Figure \ref{fig.3He}, the experimental data deviate from $\kappa\propto T^{-1}$ (or $WT\propto T^2$), a feature that is often taken for granted. With the exception of one rare case \cite{Dyugaev}, this feature did not attract any attention for decades. In 2024, it was proposed \cite{Behnia2024} that collective transport by a sound mode is the driver of the deviation from the quasiparticle picture of heat transport in $^3$He. 

A quantum liquid becomes a classical liquid above its degeneracy temperature. In the case of $^3$He, the ambient-pressure melting temperature and the degeneracy temperature are so close to each other that the classical liquid has not been explored at all. Conceptually, however, one can perform a thought experiment. What would be the thermal conductivity of a classical liquid composed of $^3$He atoms? There is no reason to doubt that the Bridgman equation would hold for this hypothetical liquid. Now, what would happen to the thermal conductivity of such a liquid as the temperature is reduced below $T_F$? 

In Ref. \cite{Behnia2024}, it was first shown that a Landauer-like expression for the entropy conductivity of quasiparticles, which follows a $T^{-2}$ dependence, can be written as 

\begin{equation}
\frac{\kappa}{T}|_{qp}=\frac{2\pi}{9}\frac{k_B^2}{h} k_F^2 \ell
\label{qp}
\end{equation}

This can be viewed as a Drude conductivity in disguise, combined with the Wiedemann-Franz law. $\frac{k_B^2}{h}$ is the quantum of entropy conductance. The number of conducting channels is proportional to $k_F^2$ and $\ell\propto T^{-2}$ is the quasiparticle mean free path. Note that this quadratic temperature dependence arises from the variation of the average distance between two collisions. It is more appropriate to attribute this behavior to a `quasiparticle gas' rather than to a `Fermi liquid'. As discussed in the previous section, in liquids transport is collective, in qualitative contrast to gases.

If collective transport in the quantum liquid involves transmission by the whole Fermi surface, then one may replace the mean free path in Eq.~\ref{qp} with the de Broglie thermal length, $\Lambda = \frac{h}{\sqrt{2\pi m^* k_B T}}$ where $m^*$ is the effective mass, which quantifies the thermal fuzziness of the Fermi surface \cite{Behnia2015b}). This leads to \cite{Behnia2024}:

\begin{equation}
\frac{\kappa}{T}|_{s}=\frac{2\pi}{9}\frac{k_B^2}{h} k_F^2 \Lambda
\label{kappa_zs}
\end{equation}

Substituting the Fermi wave vector using the particle density ($n=\frac{k_F^3}{3 \pi^2}$) and replacing $\Lambda$ with its explicit expression, this expression becomes:

\begin{equation}
\kappa_{s}= r k_B n^{2/3}\sqrt{\frac{2 k_B T}{m^*}}
\label{kappa_zs2}
\end{equation}

In this equation, which is to be compared with Eq.~\ref{Bridgman}, $r=\frac{\pi^{11/6}}{ 3^{4/3}}\approx 1.88$. The two equations are equivalent under the corresponding substitutions. Eq.~\ref{Bridgman} becomes Eq.~\ref{kappa_zs2} after replacing $rk_B$ by k$_B$T/$T_F$, v$_s$ by $v_F$, and $n^{-1/3}$ by $\Lambda$.

Assuming such an additional conduction channel, the experimental data were fitted over the temperature range from 0.01 K to 3 K, using \cite{Behnia2024}:

\begin{equation}
\kappa(T)= aT^{-1}+bT^{1/2}
\label{total}
\end{equation}

The first term represents the quasiparticle contribution, $\kappa_{qp} \propto T^{-1}$, and the second the collective (sound) contribution, $\kappa_s \propto T^{1/2}$.

It is instructive to rewrite Eq.~\ref{qp} in another fashion. Let us replace the mean free path by the scattering time: $\ell =v_F \tau$. The latter can be written \cite{Gourgout2024}: 

\begin{equation}
\tau= \zeta^{-1} \frac{\hbar E_F}{(k_BT)^2}
\label{zeta}
\end{equation}

The dimensionless parameter $\zeta$, defined by this equation, quantifies the strength of the fermion--fermion collisions. It is intimately linked with the Landau parameters of the Fermi liquid. Combining Eqs.~\ref{qp} and \ref{kappa_zs} yields:

\begin{equation}
\frac{\kappa_s}{\kappa_{qp}}=\pi\zeta\left(\frac{T}{T_F}\right)^{3/2}
\label{ratio}
\end{equation}

\begin{figure*}[ht!]
\centering
\includegraphics[width=1\linewidth]{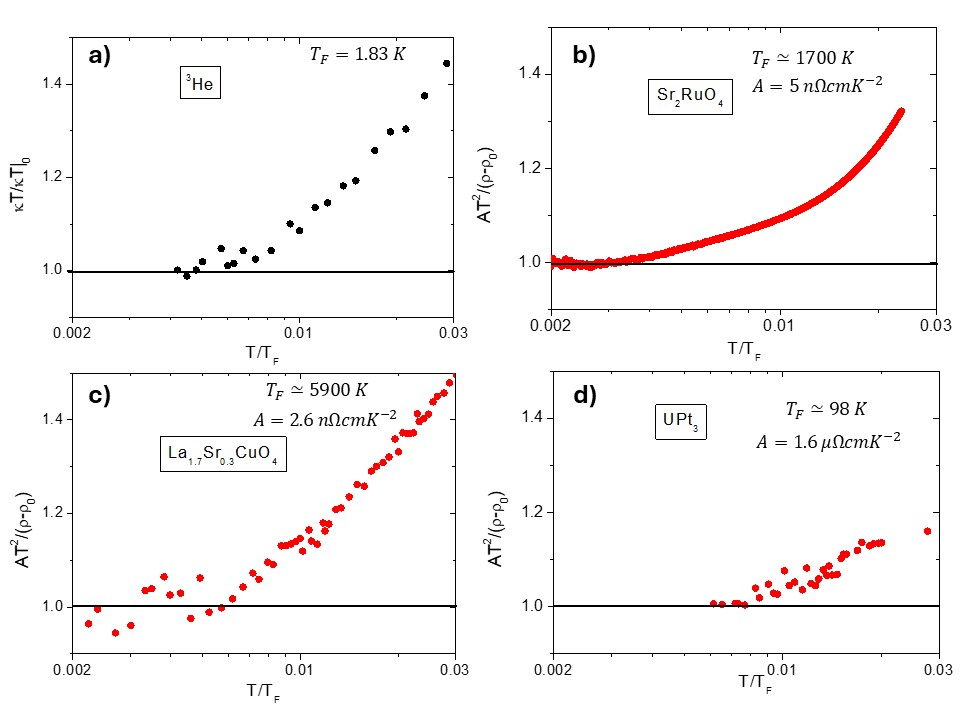} 
\caption{\textbf{Correction to the quadratic conductivity in $^3$He and in metals} (a) Thermal conductivity multiplied by temperature, $\kappa T$, in $^3$He \cite{greywall1984} divided by its zero-temperature extrapolation as a function of reduced temperature ($T/T_F$). The excess conductivity at $T/T_F=0.003$ is about 45 percent. (b) to (d) Temperature-dependent electrical conductivity $(\rho-\rho_0)^{-1}$ multiplied by $AT^2$ as a function of reduced temperature in three strongly correlated metals plotted with the same horizontal and vertical scales. In the case of a strict $T^2$ dependence, one would have a flat data set. The observed excess of conductivity at $T/T_F=0.03$ is to be compared with what is found in $^3$He.}
\label{fig.comp}
\end{figure*}

Thus, the sound component becomes increasingly prominent with increasing temperature. When $\zeta$ is sufficiently large, the collective contribution is significant even when $\frac{T}{T_F} \ll 1$. 

This is the case of $^3$He, even at ambient pressure. The amplified effective mass is $m^\star_3=2.80m_3$ \cite{Greywall1986} and the Fermi velocity $v_F=60\,\mathrm{m\,s^{-1}}$, yielding a Fermi temperature $T_F=1.83$ K. Thermal conductivity data yield $\tau T^2=0.391\times10^{-12}\,\mathrm{s\,K^2}$. Substituting these numbers into Eq.~\ref{zeta} yields $\zeta\simeq35$ \cite{Behnia2022b}. 

Note that the value of $\zeta$ can be calculated from the tabulated values of the Landau parameters of $^3$He \cite{wolfle1979} (see the Appendix). As a consequence of this large $\zeta$, in $^3$He, the two contributions become equal when $T/T_F = 0.064$. Indeed the fit to the data using Eq.~\ref{total} finds that at a temperature as low as $T\simeq0.1$ K, half of the total conductivity is carried by the sound component. 


Thus, despite not being rigorously derived, the conceptual plausibility and the empirical success of Eq.~\ref{kappa_zs} suffice for taking it seriously as the seed of an explanation for transport in a Fermi liquid at finite temperatures. 

\section{Comparison of excess conductivity in four Fermi liquids}

Let us return to the question raised at the beginning of this paper. What causes the downward deviation from $T^2$ resistivity in metallic Fermi liquids? It has been implicitly assumed that the deviation is due to the presence of an additional energy scale much smaller than the Fermi energy.

An empirical comparison between the excess conductivity in $^3$He and in metallic Fermi liquids would be instructive. There is one obstacle, however. In a multiband metal, there are multiple Fermi energies. 

UPt$_3$ has five partially filled bands, all detected by quantum oscillations \cite{McMullan_2008}. The detected de Haas--van Alphen (dHvA) masses and frequencies imply a variety of Fermi energies for different pockets, which have been dubbed Starfish ($\approx$ 322 K), Octopus ($\approx$ 65 K), Oyster and Urchins ($\approx 95$ K), Mussel ($\approx$ 181 K) and Pearl ($\approx$ 14 K) \cite{McMullan_2008}. To simplify the comparison, we have opted for an average Fermi energy of 98 K, which has been extracted from its Sommerfeld coefficient by assuming a single electron per unit cell. 

Sr$_2$RuO$_4$ has three quasi-two-dimensional Fermi surface pockets \cite{Mackenzie2003}. Computing their Fermi energies, using $E_F=\hbar^2k_F^2/2m^\star$ from the measured $k_F$ and $m^\star$ values, yields three Fermi energies ranging from 1200 K to 2400 K. We have opted for an average Fermi energy of 1700 K. 
La$_{1.7}$Sr$_{0.3}$CuO$_4$ has a single Fermi pocket and its Fermi temperature has been estimated by several converging approaches to be 5900 K \cite{Behnia2022,Jin2021}.

\begin{table*}[htbp]
\centering
\caption{\textbf{Fermi-liquid details of four systems} 
The average Fermi energy, the average Fermi wave-vector, and the prefactor of the $T^2$ resistivity, $A$, were injected to either
$\zeta_{3d} = \frac{2 e^2 }{3 \pi^2 \hbar}k_F E_F^2 A$ or $\zeta_{2d} = \frac{e^2 }{3 \pi c \hbar}k_F E_F^2 A$. The extracted values of $\zeta$ are listed.}
\label{tab:comparison}
\begin{tabular}{ccccc}
\hline
System & $T_{F}$(K) & $k_F$ (nm$^{-1}$) & A ($\mu \Omega$cm\,K$^{-2}$) &$\zeta$ \\
\hline
\hline
$^3$He & 1.83 & 7.9 &-- &35 \\

\hline
UPt$_3$ & 98 & 7 & 1.6 & $\approx$ 17 \\
\hline
Sr$_2$RuO$_4$ & 1700 & 6.5 & 0.005&$\approx$16 \\
\hline
La$_{1.7}$Sr$_{0.3}$CuO$_4$& 5900 &5.6& 0.0026 & 24-61 \\
\hline
\end{tabular}
\end{table*}

Replacing multiple Fermi energies by a single average value in multiband metals is a simplification that should be kept in mind in what follows. It allows a comparison of the relative amplitudes of the deviations. Figure \ref{fig.comp}(a) shows $\kappa T/ \kappa T|_0$ in $^3$He. $\kappa T|_0$ is the zero-temperature extrapolated value \cite{greywall1984}.  In the quasiparticle picture, this ratio would have remained flat. The experimental data reveal a deviation starting at a temperature that is orders of magnitude smaller than the average [or even the lowest] Fermi temperature of the system. 

The other panels of Figure \ref{fig.comp} display the temperature dependence of $AT^2(\rho-\rho_0)^{-1}$ in the same window of reduced temperature in three strongly correlated metals (Sr$_2$RuO$_4$ \cite{Barber2018}, La$_{1.7}$Sr$_{0.3}$CuO$_4$ \cite{Nakamae2003} and UPt$_3$ \cite{Kimura}). It is striking to see that the relative excess of conductivity at a reduced temperature of $T/T_F=0.02$ in Sr$_2$RuO$_4$ and in La$_{1.7}$Sr$_{0.3}$CuO$_4$ is comparable to what is found in $^3$He. In UPt$_3$, it is only half as large. Thus, when the temperature is still 50 times below the [average] Fermi temperature, the deviation from the zero-temperature extrapolated value is between 18 and 40 percent in these four cases.

It is tempting to attribute the excess conduction to a second conducting channel in metallic Fermi liquids, as was postulated in $^3$He \cite{Behnia2024}. Collective modes in strongly correlated Fermi liquids have been the subject of several studies \cite{Dornheim2016,Takada2016}. To the best of our knowledge, however, their role as a conducting channel has never been discussed. 

The interpretation proposed here is not the only available explanation for the observed deviation from the $T^2$ resistivity. The existence of energy scales other than the Fermi energy in many strongly correlated metals can provide an alternative explanation. In heavy-fermion systems, the Kondo lattice has a characteristic energy scale \cite{LACROIX1986145,Assaad_2004}. However, this energy scale is not expected to be as low as $0.02T_F$. In the specific case of UPt$_3$, the coherence temperature has been experimentally detected to be $T_{coh}\simeq 30$ K \cite{Dressel2002}. This is one-third of what we took as the average Fermi energy of the system, but much larger than the $\simeq$ 1 K temperature scale at which the deviation from quadratic temperature dependence is detectable. In ruthenates and pnictides, the strict $T^2$ regime has been suggested to be bounded by a coherence-incoherence crossover temperature. The latter can be as low as 50 K \cite{Mravlje2011} in Sr$_2$RuO$_4$. This is thirty times lower than what we took as the average Fermi temperature but significantly larger than the onset of the deviation from the quadratic behavior. 

One cannot exclude the existence of additional energy scales and their possible role in the observed deviation. Therefore a transport picture exclusively based on quasiparticles cannot be definitely ruled out. However, the phenomenological similarity of Fermi liquids, which, despite their different absolute energy scales and microscopic details, have a similar excess of conductivity at similar $T/T_F$ and a two-digit value of $\zeta$ (Table I), argues in favor of the present scenario, without establishing it definitively.

A question remains. Can a charge-neutral collective mode carry charge, as well as heat? This is an open question. We know, however, that sliding density waves, which are also charge-neutral, can respond to an electric field and contribute to electrical transport \cite{Ong1977}.

In summary, we report an observation and a speculation. Strongly correlated Fermi liquids consistently show a downward deviation from the canonical $T^2$ temperature dependence well below the Fermi degeneracy temperature. This may be caused by the presence of a collective sound-like channel of conduction with a distinct temperature dependence. The weight of this collective mode is expected to be larger in more strongly correlated metals. This hypothesis accounts for the temperature dependence of thermal conductivity in $^3$He over a broad temperature range.

\section{Acknowledgments}
The author is grateful to Andy Mackenzie and Mark Barber for sharing their resistivity data of Sr$_2$RuO$_4$ \cite{Barber2018}.

\bibliography{biblio}
\newpage
\section{Appendix}
In a Fermi liquid, the available phase space for quasiparticle scattering scales quadratically with temperature $T$. Consequently, the quasiparticle collision rate $\tau^{-1}$ obeys Eq.~\ref{zeta} of the main text. This defines the dimensionless parameter $\zeta$:

\begin{equation}
 \zeta = \tau^{-1} \frac{\hbar E_F}{(k_B T)^2}
\end{equation}
where $E_F = \frac{1}{2} m_3^\star v_F^2 = k_B T_F$ is the Fermi energy, $T_F$ is the Fermi temperature, and $\zeta$ is a dimensionless parameter characterizing the angle-averaged scattering probability over the Fermi surface.

In normal liquid $^3\text{He}$ at $p=0\,\mathrm{bar}$, according to thermal conductivity measurements, in the zero-temperature limit 
$\tau T^2 \rightarrow 0.391\,\mathrm{ps\,K^2} $ \cite{greywall1984}. The combination of the Fermi velocity and the effective mass derived from specific heat \cite{Greywall1986} yields $T_F \simeq 1.83 \text{ K}$. This yields an experimental value of \cite{Behnia2022b}:
\begin{equation}
\zeta_{\text{exp}} = \frac{\hbar T_F}{k_B (\tau T^2)} \approx 35.7
\end{equation}

The magnitude of this two-digit value of $\zeta$ can be traced to the known Landau parameters of $^3$He tabulated by Vollhardt and Wolfle \cite{vollhardt2013superfluid}. For detailed calculations of the quasiparticle scattering cross section, see \cite{wolfle1979,Pfitzner1985,Pfitzner1987}.

The standard Fermi-liquid transport theory \cite{Dy1969,Brooker1968} relates the collision rate parameter $\zeta$ to the phase-space integral of $W(\theta, \phi)$:
\begin{equation}
\zeta = \frac{\pi^2}{8} \langle W \rangle
\end{equation}
where the Fermi surface weighted phase-space average $\langle W \rangle$ is defined as:
\begin{equation}
\langle W \rangle = \int_0^{\pi} \frac{\sin\theta \, d\theta}{2} \int_0^{2\pi} \frac{d\phi}{2\pi} \, \frac{W(\theta, \phi)}{2 \cos(\theta/2)}
\end{equation}
The factor $(2\cos(\theta/2))^{-1}$ represents the density of final phase-space states available for collisions on the Fermi sphere at angle $\theta$. $W(\theta, \phi)$ is a spin-averaged dimensionless transition probability:
\begin{equation}
W(\theta, \phi) = \frac{1}{4} \left| A_{\text{singlet}}(\theta, \phi) \right|^2 + \frac{3}{4} \left| A_{\text{triplet}}(\theta, \phi) \right|^2
\end{equation}

Scattering occurs in spin-singlet or spin-triplet configurations. The spin combinations of the scattering amplitudes are:
\begin{align}
A_{\text{singlet}}(\theta, \phi) &= A^s(\theta, \phi) - 3 A^a(\theta, \phi) \\
A_{\text{triplet}}(\theta, \phi) &= A^s(\theta, \phi) + A^a(\theta, \phi)
\end{align}

Truncating at $l = 1$ ($s$-wave and $p$-wave channels), the full scattering amplitudes $A^s$ and $A^a$ as functions of scattering angles are:
\begin{align}
A^s(\theta, \phi) &= A_0^s + A_1^s \cos\theta_{\text{scat}} \\
A^a(\theta, \phi) &= A_0^a + A_1^a \cos\theta_{\text{scat}}
\end{align}

Here, $\theta_{\text{scat}}$ is the scattering angle between the initial and final quasiparticle momenta:
\begin{equation}
\cos\theta_{\text{scat}} = \cos^2\left(\frac{\theta}{2}\right) + \sin^2\left(\frac{\theta}{2}\right)\cos\phi
\end{equation}

The effective scattering amplitudes $A_l^{s,a}$ are related to the Landau interaction parameters $F_l^{s,a}$ via:
\begin{equation}
A_l^s = \frac{F_l^s}{1 + \frac{F_l^s}{2l + 1}}, \qquad A_l^a = \frac{F_l^a}{1 + \frac{F_l^a}{2l + 1}}
\end{equation}

Substituting $F_0^s \approx 9.3$, $F_1^s \approx 5.4$, $F_0^a \approx -0.7$, and $F_1^a \approx -0.55$ \cite{vollhardt2013superfluid}, leads to $A_0^s \approx 0.903; A_1^s \approx 1.93; A_0^a \approx -2.33$, and $A_1^a \approx -0.673$. With these numbers, the double integral gives $\langle W \rangle \approx 23.82$ and the resulting value of $\zeta$ is:
\begin{equation}
\zeta_{\text{calc}} = \frac{\pi^2}{8} \times 23.82\approx 29.4
\end{equation}

Given that we have neglected all Landau parameters with $\ell>1$, this is remarkably close to $\zeta \approx 35$. Note that $\zeta$ is highly sensitive to the amplitude of $F_0^a$, which quantifies the proximity of the system to a magnetic instability. As an example, assuming $F_0^a=-0.72$ leads to $\zeta=35$.

\end{document}